\documentclass[12pt,a4,epsf]{article}
\global\arraycolsep=2pt 
\input{epsf}
\usepackage{graphicx}
\usepackage{amssymb}
\usepackage{cite}

\def\eq#1\en{\begin{equation}#1\end{equation}}
\def\s[#1,#2]{[#1\stackrel{\star}{,}#2]}

\begin{document}
\makeatletter
\def\fmslash{\@ifnextchar[{\fmsl@sh}{\fmsl@sh[0mu]}}
\def\fmsl@sh[#1]#2{%
  \mathchoice
    {\@fmsl@sh\displaystyle{#1}{#2}}%
    {\@fmsl@sh\textstyle{#1}{#2}}%
    {\@fmsl@sh\scriptstyle{#1}{#2}}%
    {\@fmsl@sh\scriptscriptstyle{#1}{#2}}}
\def\@fmsl@sh#1#2#3{\m@th\ooalign{$\hfil#1\mkern#2/\hfil$\crcr$#1#3$}}
\makeatother
\thispagestyle{empty}
\begin{titlepage}
\begin{flushright}

hep-ph/0302056 \\
CALT-68-2427\\
March 17, 2003
\end{flushright}

\vspace{0.3cm}
\boldmath
\begin{center}
  \Large {\bf Softening the  Naturalness Problem}
\end{center}
\unboldmath
\vspace{0.8cm}

\unboldmath
\vspace{0.8cm}
\begin{center}
  {\large Xavier Calmet \footnote{
email:calmet@theory.caltech.edu}}\\
\end{center}
 \vspace{.3cm}
\begin{center}
{\sl California Institute of Technology, Pasadena, California
91125, USA}
\end{center}
\vspace{\fill}
\begin{abstract}
\noindent
 
\end{abstract}
It was observed by Veltman a long time ago that a special value for
the Higgs boson mass could lead to a cancellation of the quadratically
divergent corrections to the Higgs boson's squared mass which appear
at one loop. We present a class of low energy models that allow to
soften the naturalness problem in the sense that there can be a
cancellation of radiative corrections appearing at one loop. The
naturalness problem is shifted from the one TeV region to the 10 TeV
region. Depending on the specific model under consideration, this
scale can even be shifted to a higher energy scale. Signatures of
these models are discussed.
\end{titlepage}

\section{Introduction}
The standard model is a gauge theory based on the structure group
 SU(3) $\times$ SU(2) $\times$ U(1)
 \cite{Glashow:1961tr,Fritzsch:2002jv}.  The electroweak gauge
 symmetry is spontaneously broken by means of the Higgs mechanism
 \cite{Higgs:1964pj}. This mechanism requires the inclusion in the
 model of a fundamental scalar field which is charged under U(1) and
 is in the fundamental representation of SU(2). It is often argued
 that the standard model with the Higgs mechanism as a mechanism for
 gauge symmetry breaking cannot be a theory valid over a wide range of
 energies because the squared mass of a scalar field receives
 corrections that are quadratic divergent if a naive cutoff is used to
 regularize the model. In the standard model one obtains
\begin{eqnarray} \label{Veltmanori}
m_H^2 \approx {m^0_H}^2 + \frac{3 g^2 \Lambda^2}{32 \pi^2 m_W^2}
\left( m_H^2 + 2 m_W^2 + m_Z^2 -4 \sum_f \left( \frac{n_f}{3}\right)
m_f^2\right) 
\end{eqnarray} 
in the one loop approximation \cite{Veltman:1981mj}. The mass term of
the Higgs boson is denoted by $m_H$, that of the W-boson by $m_W$,
that of the Z-boson by $m_Z$ and that of the top-quark is denoted by
$m_t$. Finally $n_f$ is the number of flavors propagating in the loop
and $g$ is the SU(2) gauge coupling constant. If the cutoff is large,
e.g. of the order of a possible grand unification scale $10^{16}$ GeV,
it requires an unnatural adjustment between the bare mass $m_H^0$ of
the scalar field and the ``corrections'' to insure a physical Higgs
mass in the 100 GeV region \cite{'tHooft:1980xbis}. This is known as
the naturalness problem. It should nevertheless be noted that the
standard model is renormalizable \cite{'tHooft:rn}. Quadratic
divergences can be absorbed in the parameters of the standard model in
a way which is mathematically completely consistent. 

An obvious solution to the naturalness problem is to avoid the Higgs
mechanism and the inclusion of fundamental scalars in the model like
in e.g. technicolor theories or top condensate models
\cite{Hill:2002ap}. Another approach is to embed the standard model
into another more fundamental theory where quadratic divergences are
either absent like in supersymmetric models \cite{Haber:1984rc} or
small like in models with extra-dimensions \cite{Arkani-Hamed:1998rs}
because in that case the fundamental scale of nature is assumed to be
not much larger than the electroweak scale.

Recently another point of view has been revived. It had been proposed
a long time ago \cite{Georgi:1975tz}, that the Higgs boson could be a
pseudo-Goldstone boson. This idea has recently been revived in the
form of the Little Higgs models \cite{Arkani-Hamed:2002qx}. The Little
Higgs models have been shown to possess an approximate symmetry that
can protect the Higgs mass from radiative corrections if the cutoff is
not too large i.e. 10 TeV. Unfortunately the simplest examples of that
class of models do not automatically have a custodial symmetry and are
thus potentially severely constrained by experiments
\cite{Csaki:2002qg}.

Long before that, it had been speculated that a special value for the
Higgs boson mass could cancel the quadratic divergences corrections to
the Higgs boson mass \cite{Veltman:1981mj}. This leads to the
so-called Veltman's relation:
\begin{eqnarray}
m_H^2=4 m_t^2-2 m_W^2-m_Z^2.
\end{eqnarray}
But, even if this relation was fulfilled, i.e. if the Higgs boson mass
was of the order of $316$ GeV, it does not hold beyond the one loop
approximation. The consequences of a possible cancellation of the
logarithmic divergent terms have been considered in
\cite{Decker:gw}. Note that there is no need to require an exact
cancellation of the one loop quadratic divergences
\cite{Kolda:2000wi}.

In this work we shall argue that the observation made by Veltman some
22 years ago can be revived in a modern framework. We shall present a
class of models for which, as in the Little Higgs models case, we do
not describe the high energy completion and we shall thus assume that
our models have a cutoff in the 10 TeV region. The simplest way to do
that is to assume that the standard model Higgs boson mass is of the
order of 316 GeV and that this model has a cutoff of 10 TeV. We note
that such a high mass for the Higgs boson seems to be in contradiction
with fits based on electroweak precision measurements
\cite{Hagiwara:fs}. Furthermore, in that case, there is no way to push
the cutoff scale above the 10 TeV scale. Our basic observation is that
if there is some new physics beyond the standard model with a new
bosonic degree of freedom $\phi$ that is coupled to the standard model
in a minimal way i.e.
\begin{eqnarray} \label{newop}
\alpha h^\dagger h \phi^\dagger \phi,
\label{newterm}
\end{eqnarray} 
where $h$ is the standard model Higgs boson and $\alpha$ is a
parameter of order one, then the Higgs boson mass can naturally be of
the order of 100 GeV if the models have a cutoff of the order of 10
TeV. Furthermore, depending on the model under consideration, this
cutoff scale can be pushed upwards. The naturalness problem is not
solved by these models, but is only soften. It must be emphasized that
these models are theoretically not as compelling as the Little Higgs
models, because they do not have an approximate symmetry that protects
the Higgs boson squared mass against radiative corrections but they
have a custodial symmetry and are thus not constrained by
experiments. These models are semi-natural in sense proposed by
Veltman \cite{Veltman:1981mj}. A new boson that couples to the Higgs
boson of the standard model according to (\ref{newterm}) implies a
correction to the Higgs boson squared mass given by:
\begin{eqnarray}
m_H^2&\approx& {m^0}_H^2+ \alpha \Lambda_c^2 
\\ \nonumber &&
+ \frac{3 g^2 \Lambda^2}{32 \pi^2 m_W^2}
\left( 4 v^2 \lambda + 2 m_W^2 + m_Z^2 -4 \sum_f \left( \frac{n_f}{3}\right)
m_f^2\right) + \frac{\Lambda^2}{4 \pi^2} \alpha ,
\end{eqnarray}
where $\lambda$ is the Higgs boson self-coupling, $v=174$ GeV the
vacuum expectation value and $\Lambda_c$ is the scale associated with
the new physics beyond the standard model.  Note that depending on the
model our definition of $\Lambda_c$ includes a potential mixing angle
between the scales of the different scalar sectors.  In the sequel we
shall describe two classes of models that should be considered as low
energy effective theories of an unknown high energy theory. If we
introduce the operator (\ref{newop}) in the standard model, we
potentially introduce a new naturalness problem. Because $\phi$ is a
bosonic degree of freedom, its squared mass will in general receive
quadratic corrections. We have identified two classes of models where
this problem is under control.  The first class of models are models
where $\Lambda_c$ is identical to the standard model electroweak
scale, and where a symmetry implies that the corrections to the
squared mass of the scalar field $\phi$ are identical to those of the
scalar fields $h$. The second class of models are models where the
scale involving the second scalar field $\phi$, i.e. its mass, is
rather near to the cutoff scale $\Lambda$. The corrections to its
squared mass are thus small.

Clearly, if $\alpha$ is a positive parameter of the order of one, a
cancellation, or partial cancellation, of the the quadratic
corrections can take place and the Higgs boson mass can naturally be
of the order of 100 GeV if the cutoff is of the order of 10 TeV. Note
that the the two loop corrections are expected to be of the order of
$(\frac{ 1 }{16 \pi^2})^2 \Lambda^2$ and are thus small if the cutoff
is as low as 10 TeV. The corrections to the potential of the new
scalar degree of freedom are model dependent and shall be discussed
below for each model considered. Note also that if $\alpha$ is a
negative parameter of order one and if $\Lambda_c< \Lambda$, there is
a negative contribution at tree level to the Higgs boson mass that
can, in principle, reverse the sign of the Higgs boson squared mass
and thus trigger the Higgs mechanism in the electroweak sector of the
standard model. In that scenario we have to require that the new scale
is lower than the cutoff scale to be certain that the low energy
effective theory remains valid. Nevertheless, in that case, there is,
in general, no cancellation of the quadratic divergences.

In the first section, we shall consider a model where $\Lambda_c$ is
assumed to be equal to the scale of the electroweak interaction. In
the second section we will describe a model where mass of the second
scalar field is assumed to be near to the cutoff of the models. We then
conclude.

\section{A standard model replica}

Let us consider a model based on the gauge group SU(3)$_{\bar C}$
$\times$ SU(2)$_{\bar L}$ $\times$ U(1)$_{\bar Y}$ $\times$ SU(3)$_C$
$\times$ SU(2)$_L \times$ U(1)$_Y$, where SU(3)$_{\bar C}$ $\times$
SU(2)$_{\bar L}$ $\times$ U(1)$_{\bar Y}$=G$_n$ is the gauge group
describing the physics beyond the standard model and SU(3)$_C \times$
SU(2)$_L \times$ U(1)$_Y$=G$_{SM}$ is the standard model gauge
group. Both SU(2) groups have the usual weak gauge coupling $g$ and
both SU(3) groups have the usual strong coupling $g_s$. Furthermore,
we assume also that both U(1) groups have the same usual gauge
coupling $g'$. The number of fields of the model is thus doubled in
comparison to the standard model (see table \ref{tab:table1}), and
there is a discrete symmetry transforming the standard model fields
(Higgs boson included) into the fields charged under G$_n$.  The
motivation to consider this model is that because of the discrete
symmetry, the corrections to the masses of both scalar degrees of
freedom are identical. Therefore only one relation has to be fulfilled
to insure the naturalness of the model if the cutoff is assumed to be
around 10 TeV. The Lagrangian of the model is given by
\begin{eqnarray}
{\cal L}_{rep}={\cal L}_{SM}+{\cal L}_{n}-\alpha h^\dagger h \Phi^\dagger \Phi
\end{eqnarray}
where $h$ is the SU(2)$_L$ scalar doublet and $\Phi$ is the
SU(2)$_{\bar L}$ scalar doublet. This is the most generic, gauge
invariant Lagrangian. ${\cal L}_{SM}$ is the standard model Lagrangian
and ${\cal L}_n$ is the Lagrangian containing the new fields, it is
obtained by applying the discrete symmetry mentioned above on ${\cal
L}_{SM}$. Note that this model has a custodial symmetry and is thus
not in contradiction with electroweak precision measurements.  The
potential of the model reads:
\begin{eqnarray}
V(h,\Phi)&=& m_s^2 h^\dagger h 
           +m_s^2 \phi^\dagger \phi 
	  +\lambda (h^\dagger h)^2 
           +\lambda (\Phi^\dagger \Phi)^2 
           + \alpha h^\dagger h \Phi^\dagger \Phi.
\end{eqnarray}
The vacuum expectation value of the scalar fields $v$
is given by:
\begin{eqnarray}
v&=&\sqrt \frac{-m^2_s}{2 \lambda+\alpha},
\end{eqnarray}
note that $m_s^2<0$, the potential is bounded from below if $2
\lambda>\alpha$. The parameter $m_s$ receives the usual quadratic
divergencies:
\begin{eqnarray}
m_s^2 &\approx& {m^0}_s^2+\alpha v^2 
\\ \nonumber &&
+ \frac{3 g^2 \Lambda^2}{32 \pi^2 m_W^2}
\left( 4 v^2 \lambda + 2 m_W^2 + m_Z^2 -4 \sum_f \left( \frac{n_f}{3}\right)
m_f^2\right) + \frac{ \Lambda^2}{4 \pi^2} \alpha.
\end{eqnarray}
If we require the complete cancellation of the quadratic divergent
corrections induced at one loop, we obtain the analog of Veltman's
condition:
\begin{eqnarray}
\lambda+ \frac{1}{3}\alpha
=\frac{g^2}{8 m_W^2}\left(4 \sum_f \left( \frac{n_f}{3}\right)
m_f^2- 2 m_W^2 - m_Z^2\right)\approx 0.86,
\end{eqnarray}
which insures the cancellation of the quadratic divergences induced by
the one loop corrections.

{\bf Signatures} 

The operator $h^\dagger h \Phi^\dagger \Phi$ induces a mixing between
the two scalar doublets. Let us denote the mass eigenstates by $h_1$
and $h_2$. One finds:
\begin{eqnarray}
h_1&=&\frac{1}{\sqrt{2}}h^0- \frac{1}{\sqrt{2}}\phi^0 \\  \nonumber
h_2&=&\frac{1}{\sqrt{2}}h^0+ \frac{1}{\sqrt{2}}\phi^0, 
\end{eqnarray}
i.e. the mixing is maximal. After this diagonalization procedure,
$h_1$ and $h_2$ couple to both the fermions and their replicas. There
is nevertheless no new source of neutral flavor changing. The replicas
are only extremely weakly coupled to the standard model particles, the
only possible interaction is mediated by the scalar bosons. In that
sense the decays of the scalar bosons to replicas should be considered
as missing energy decay modes. Therefore, the observable spectrum of
the theory is the standard model with a further neutral scalar
boson. The squared masses of the scalar bosons are given by
\begin{eqnarray}
m^2_{h_1}&=&{m^0}_h^2-\alpha v^2=2(2 \lambda -\alpha) v^2 \\ \nonumber
m^2_{h_2}&=&{m^0}_h^2+3\alpha v^2= 2(2 \lambda +\alpha) v^2,
\end{eqnarray}
with $v=174$ GeV. Let us assume that the lightest Higgs boson has a
mass $m_{h_1}$ of about 130 GeV and that the one loop quadratically
divergent corrections cancel completely, one finds $m_{h_2}=349$
GeV. Note that the production and decay modes of these scalar bosons
are quite different from the standard model case. At an electron
positron collider the production rate is $1/2$ of that of the standard
model because of the mixing between the two scalars reduces the
coupling of each of the scalars to the SU(2)$_L$ gauge bosons by a
factor $1/\sqrt{2}$. One has schematically $\sigma(e^+ e^- \to H Z^*)
\sim \frac{1}{2} \sigma(e^+ e^- \to h_1/h_2 Z^*)$ where is it
understood that the appropriate scalar mass has to be used in the
formula. Similarly the couplings to fermions charged under G$_{SM}$,
i.e. the standard model fermions, is reduced by the same factor. This
implies that the LEP limits for these scalar bosons are less stringent
in this model. The lightest Higgs boson decays as in the standard
model dominantly to $b$-quarks if its mass is around 100 GeV. The
cross section is $\sigma(e^+ e^- \to b \bar b Z^*)= \left.\frac{1}{4}
\sigma(e^+ e^- \to b \bar b Z^*)\right|_{SM}$, and is thus much
smaller than in the standard model. We have assumed that only the
lightest Higgs boson contributes at a significant level. The cross
section to missing energy, i.e. when the Higgs boson decays to the
$b$-quark replicas, is equal in magnitude to the $b$-$\bar b$ decay
mode: $\sigma(e^+ e^- \to \mbox{missing energy} \ Z^*)=\sigma(e^+ e^-
\to b \bar b Z^*)$. The missing energy corresponds to a scalar of mass
$m_{h_1}$. Notice that both scalars couple both to the standard model
particles and to the replica particles. This implies that fifty
percent of the scalar bosons should be missing energy decays. Note
that the production at a hadron collider, where one of the main
production mode for the Higgs boson is via a top quark triangle, will
also be suppressed by a factor $1/2$ compared to the standard model
expectation.

\begin{table}
\centering
  \begin{tabular}{|c|c|c|c|c|c|c|}
  \hline
   &SU(3)$_C$& SU(2)$_L$ & U(1)$_Y$  &
  SU(3)$_{\bar C}$ & SU(2)$_{\bar L}$ & U(1)$_{\bar L}$
   \\
   \hline
     $ e_R$
& ${\bf 1}$ 
   & ${\bf 1}$ 
   & $-1$
& ${\bf 1}$ 
 & ${\bf 1}$
& $0$
  \\
  \hline
   $ L_L=\left(\begin{array}{c} \nu_L \\ e_L \end{array} \right )$
& ${\bf 1}$ 
   & ${\bf 2}$
   & $-1/2$
   & ${\bf 1}$
 & ${\bf 1}$
& $0$
 \\
  \hline
   $u_R$
 & ${\bf 3}$
   & ${\bf 1}$
   & $2/3$
 & ${\bf 1}$
 & ${\bf 1}$
& $0$
  \\
  \hline
   $d_R$
 & ${\bf 3}$
   & ${\bf 1}$
   & $-1/3$
& ${\bf 1}$
 & ${\bf 1}$
& $0$   
\\
  \hline
     $Q_L=\left(\begin{array}{c}  u_L \\ d_L \end{array} \right )$
& ${\bf 3}$
   & ${\bf 2}$
   & $1/6$  
 & ${\bf 1}$
& ${\bf 1}$
& $0$
  \\
\hline
$h=\left(\begin{array}{c}  h^+ \\  h^0 \end{array} \right )$
  & ${\bf 1}$ 
& ${\bf 2}$ 
   & $1/2$
& ${\bf 1}$
   & ${\bf 1}$ 
   & $0$
\\
  \hline
  \hline
$ f^2_R$
& ${\bf 1}$
  & ${\bf 1}$
& $0$
& ${\bf 1}$
 & ${\bf 1}$ 
   & $-1$
  \\
\hline
   $ F_L=\left(\begin{array}{c} f^1_L \\ f^2_L \end{array} \right )$
  & ${\bf 1}$
& ${\bf 1}$
& $0$
& ${\bf 1}$
 & ${\bf 2}$
   & $-1/2$
 \\
  \hline
   $k^1_R$
& ${\bf 1}$
& ${\bf 1}$
& $0$   
& ${\bf 3}$
   & ${\bf 1}$
   & $2/3$
  \\
  \hline
   $k^2_R$
& ${\bf 1}$
& ${\bf 1}$
& $0$   
& ${\bf 3}$
   & ${\bf 1}$
   & $-1/3$
\\
  \hline
     $K_L=\left(\begin{array}{c}  k^1_L \\ k^2_L \end{array} \right )$
   & ${\bf 1}$
& ${\bf 1}$
& $0$
& ${\bf 3}$
 & ${\bf 2}$
   & $1/6$  
 \\ 
  \hline
$\Phi=\left(\begin{array}{c}  \phi^+ \\  \phi^0 \end{array} \right )$
& ${\bf 1}$
&${\bf 1}$
& $0$
& ${\bf 1}$
   & ${\bf 2}$
   & $1/2$ 
\\
  \hline
 \end{tabular}
 \caption{Particle content of the minimal SM replica model.
 \label{tab:table1}}
\end{table}

\subsection{Elusive new physics}
Let us consider the same model as described above but we shall now
assume that the SU(2)$_{\bar L}$ gauge symmetry remains unbroken. The scalar
potential is assumed to be:
\begin{eqnarray}
V(h,\Phi)&=& m_s^2 h^\dagger h 
           -m_s^2 \phi^\dagger \phi 
	  +\lambda (h^\dagger h)^2 
           +\lambda (\Phi^\dagger \Phi)^2 
           + \alpha h^\dagger h \Phi^\dagger \Phi,
\end{eqnarray}
the discrete symmetry is softly broken by the mass terms of the scalar
fields. The vacuum expectation values are given by $v_h =\sqrt
\frac{-m^2_s}{2 \lambda+\alpha}$ and $v_\Phi=0$. This implies that
there is no mixing between the two scalar fields.

The squared masses receive quadratically divergent corrections:
\begin{eqnarray}
m_h^2 &\approx& {m^0}_h^2 
\\ \nonumber &&
+ \frac{3 g^2 \Lambda^2}{32 \pi^2 m_W^2}
\left( {m}_h^2 + 2 m_W^2 + m_Z^2 -4 \sum_f \left( \frac{n_f}{3}\right)
m_f^2\right) + \frac{ \Lambda^2}{4 \pi^2} \alpha ,
\end{eqnarray}
where $m_h$ is the mass of the physical standard model Higgs boson $h$
and
\begin{eqnarray} \label{corrcopy}
m_\phi^2 &\approx& {m^0}_s^2+\alpha v^2 \\ \nonumber && + \frac{3
 \Lambda^2}{16 \pi^2} \left( 4 \lambda + \frac{3}{2} g^2 + \frac{1}{2}
 g'^2 -4 \sum_{fc} \left( \frac{n_{fc}}{3}\right)
 \lambda_{fc}^2\right) + \frac{ \Lambda^2}{4 \pi^2} \alpha,
\end{eqnarray}
where $m_\phi$ is the mass of the copy of the standard model Higgs
 boson and $\lambda_{fc}$ are the Yukawa couplings, note that we have
 $\lambda_{fc}=\lambda_{f}$. Note that if we assume that the
 parameters of the second gauge group G$_n$ are identical to those of
 G$_{SM}$ besides the sign of the mass of the scalar doublets there is
 only one relation that needs to be fulfilled:
\begin{eqnarray}
\alpha
=\frac{3 g^2}{8 m_W^2}\left(4 \sum_f \left( \frac{n_f}{3}\right)
m_f^2- 2 m_W^2 - m_Z^2 -m_h^2\right)\approx 2.13.
\end{eqnarray}
For the numerical estimate we used $m_h=130$ GeV and took only the
top quark into account. It would imply a mass of approximatively
$m_{\Phi}=\sqrt{\frac{m_H^2}{2}+\alpha v^2}\approx 270$ GeV for the
second neutral scalar boson. It is therefore natural to have a
cancellation, or near cancellation, of the quadratic divergences if
$\alpha$ is of order one. On the other hand, the corrections induced
at two loops are not vanishing. This is why we claim that the
naturalness problem is under control if there is a fundamental cutoff
for the model around 10 TeV.

{\bf Signatures}

\begin{figure}
\includegraphics[109,651][356,784]{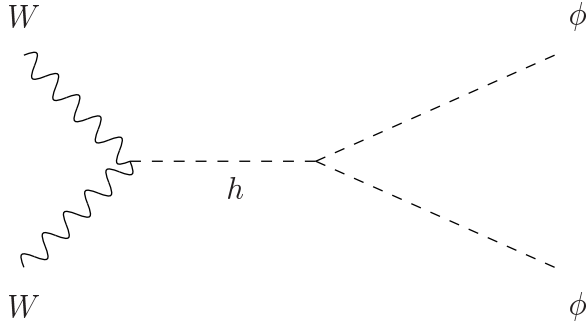}
\caption{Extraction of the Higgs bosons  coupling. The W-bosons are
emitted by the colliding fermions. \label{graph2}}
\end{figure}

A signature of this class of models is again a decay of the standard
model Higgs boson to particles that are not charged under SU(2)$_L
\times$ U(1)$_Y$. But, if only the gauge symmetry describing the gauge
interactions is broken, the new physics signals are much more
subtle. The measurement of the Higgs boson self-coupling (c.f. figure
(\ref{graph2})) becomes very interesting. In that case a sizable new
physics effect is expected. At a linear collider with a center of mass
of 1000 GeV one expects $\sigma(e^+ + e^- \to \phi \phi \bar \nu
\nu)\approx 7.4 \times 10^{-6}$ pb. This must be compared to the
standard model cross section $\sigma(e^+ + e^- \to h h \bar \nu \nu)=
8.9 \times 10^{-5}$ pb, assuming a Higgs boson mass of 130 GeV. We
have done these estimates using CompHEP \cite{Pukhov:1999gg}, taking
only the diagrams corresponding to figure (\ref{graph2}) and its
standard model counterpart into account. A large effect due to the
replica sector should be observed. This represents a further
motivation for a measurement of the Higgs boson self-coupling at a
future linear collider. The signal is a missing energy corresponding
to two scalar fields with a mass around $270$ GeV.

\subsection{A high energy completion?}

The cutoff scale can be shifted to a higher scale if we consider $N$
replicas or copies of the standard model: (SU(3)$_{\bar C} \times$
SU(2)$_{\bar L} \times$ U(1)$_{\bar Y}$)$^N \times$ SU(3)$_C \times$ SU(2)$_L
\times$ U(1)$_Y$. In our case each SU(3) $\times$ SU(2) $\times$ U(1)
group has three generations. Note that our model should not be
confused with the anti grand unification model proposed in
\cite{Bennett:1988xi}, where $N$ stands for the generation number. The
vanishing of the quadratic divergences appearing at $N$ loops requires
that $N$ equations are fulfilled. Note that if we have $N$ copies of
the standard model, we have
$N+(N-1)+(N-2)+(N-3)+(N-4)+...=\sum_{K=0}^N (N-K)$ parameters
$\lambda_N$ at our disposition. In that case the $N$ equations can be
fulfilled and the scale for the fundamental cutoff can be shifted as
high as it is necessary for model building issues.  Note that this
approach predicts numerous new particles that are stable and very
weakly interacting with the usual matter. 

\section{Second case: splitting the scales}

We now come to the second case mentioned previously, namely the case
where the scale for new physics is not far away from the cutoff scale.
In order to illustrate the idea, we shall consider a generic two Higgs
doublets model in the limit where the mass of one of the Higgs
doublets $h$ is low lying, whereas the mass of the second Higgs
doublet $H$ mass is high. The Veltman's relations for a generic two
Higgs doublets model have been considered in \cite{Newton:1993xc}. The
fermions are assumed to couple to both scalar doublets, the
hypercharge of $H$ is taken to be equal to that of $h$. This point is
however not crucial, the only requirement is that the lightest of the
Higgs bosons couples to all fermions in order to reproduce the
standard model in the decoupling limit we shall consider. In the
sequel we shall assume that the mass of the boson $H$ is rather near
to the cutoff scale. For this reason, the possible neutral flavor
changing decays are strongly suppressed. Furthermore we assume that
most of the symmetry breaking is due to the low lying doublet,
i.e. $v_1\approx v=174$ GeV and $v_2\approx 0$. In that limit the
masses of the charged and CP odd Higgs bosons are of the order of the
heavy neutral scalar boson. The assumption $v_2\approx 0$ implies that
the coefficients of the operators $h^\dagger h h^\dagger H $ or
$H^\dagger H H^\dagger h$ are nearly vanishing. Again the important
terms of the potential for our consideration are:
\begin{eqnarray}
\alpha h^\dagger h H^\dagger H + \beta h^\dagger H H^\dagger h.
\end{eqnarray}

Under the assumptions mentioned above, the leading radiative
corrections to the squared masses of the Higgs bosons are given by:
\begin{eqnarray}
m_h^2 &\approx& {m^0}_h^2
\\ \nonumber &&
+ \frac{3 g^2 \Lambda^2}{32 \pi^2 m_W^2}
\left( {m}_h^2 + 2 m_W^2 + m_Z^2 -4 \sum_f \left( \frac{n_f}{3}\right)
m_f^2\right) 
\\ \nonumber &&
+ \frac{1}{4 \pi^2} ( \alpha+\frac{1}{2}\beta) \left(\Lambda^2 - 
m_H^2 \ln\frac{\Lambda^2+m_H^2}{m_H^2} \right),
\end{eqnarray}
and 
\begin{eqnarray} \label{corrcopyB}
m_H^2 &\approx& {m^0}_H^2+(\alpha+\beta) v^2
\\ \nonumber &&
+ \frac{3 g^2}{32 \pi^2 m_{W}^2}
\left( 4 v^2 \lambda_2 + 2 m_{W}^2 + m_{Z}^2 -4 \sum_{f} \left( \frac{n_{f}}{3}\right)
m_{f}^2\right) \times 
\\
\nonumber 
&& 
 \left(\Lambda^2 - 
m_H^2 \ln\frac{\Lambda^2+m_H^2}{m_H^2} \right)
\\ \nonumber &&
 + \frac{1}{4 \pi^2} (\alpha+\frac{1}{2}\beta) \left(\Lambda^2 - 
m_H^2 \ln\frac{\Lambda^2+m_H^2}{m_H^2} \right), 
\end{eqnarray}
where $\lambda_2$ is the self-coupling of the second scalar doublet
$H$. Now we can require
the cancellation, or near cancellation, of the quadratic corrections
to the squared mass of $h$. One obtains:
\begin{eqnarray} \label{Velrel2}
\alpha+ \frac{1}{2}\beta \approx  3 \frac{3 g^2}{8 m_W^2}\left(4 \sum_f \left( \frac{n_f}{3}\right)
m_f^2-{m}_h^2 - 2 m_W^2 - m_Z^2\right)\approx 6.4,
\end{eqnarray}
using again $m_h=$ 130 GeV for the numerical estimate.  The first
factor 3 will be explained bellow. Note that although we could adjust
$\alpha$ and $\beta$ to cancel the quadratically divergent corrections
to the mass of the Higgs boson $h$, the corrections to the second
Higgs boson mass are in general rather large. This is the reason why
we assume that the mass of the second Higgs boson $m_H$ is not much
lower than the cutoff, in which case the quadratic corrections to
$m_H^2$ are small compared to its bare value. The mass of the second
scalar field $H$ could be around 9 TeV if we take a cutoff of 10 TeV.
But, if the mass of the scalar field is not much below the cutoff it
is important to consider the full one loop corrections. This explains
the term $\left(\Lambda^2 - m_H^2 \ln\frac{\Lambda^2+m_H^2}{m_H^2}
\right)$ in eq. (\ref{corrcopyB}). For $m_H=9$ TeV and $\Lambda=10$
TeV, one finds that $\Lambda^2$ should be replace by approximatively
$\Lambda^2/3$. This explains the first factor 3 in
eq. (\ref{Velrel2}).

Note that in that case, it seems very difficult to push the cutoff
scale upwards in a natural way. It has recently been pointed out that
the Higgs mechanism can be induced by a large splitting between the
two masses of a two Higgs doublets model \cite{Calmet:2002rf}. It is
thus possible to construct a two Higgs doublets model that is natural
up to a scale of 10 TeV which furthermore triggers the Higgs
mechanism.  It will be very difficult to differentiate this model from
the standard model since the low energy theory below the 9 TeV scale
is just the standard model.

We finally want to point out that it could be possible to shift the
supersymmetry breaking scale from one TeV to about 10 TeV if new
operators are added, for example, to the minimal supersymmetric
model. In that case one would have to assume that the logarithmic
terms that lead to a new naturalness problem if the supersymmetric
scale is higher than one TeV are cancelled by these new operators. We
note that supersymmetry would provide an ultraviolet completion to the
model.

\section{Conclusions}

We have described a class of models that are semi-natural in the sense
proposed by Veltman in \cite{Veltman:1981mj} a long time ago. The idea
proposed by Veltman considered in a modern framework provides an
interesting alternative to the Little Higgs models whose minimal
versions are potentially severely constrained by experiments.  In our
approach a cancellation of the radiative corrections are only
semi-natural because there is no symmetry that imposes
them. Nevertheless this is a possibility that cannot be ignored. The
new term allowing the cancellation can be generated by different types
of models. We have described two classes of models where such a term
appears.

The first type of models is a direct product of the standard model and
of its replica. If both SU(2) $\times$ U(1) symmetries are broken, the
Higgs physics is considerably affected. If only the electroweak gauge
symmetry is broken, then the new physics effects are much more subtle
and only a measurement of the scalar potential will allow to
distinguish our scenario from the standard model. This class of model
is particularly appealing since the cutoff scale can, in principal, be
shifted to any desirable scale by introducing more replicas of the
standard model.

 We then described a two Higgs doublets model where a cancellation of
the one loop quadratic divergences is possible.  The second class of
model will be much more difficult to distinguish from the standard
model and finding a deviation will require to test the very high
energy region around the mass of the second Higgs boson which can be
as high as 9 TeV.

We would like to emphasize the main point of this paper is that due to
an ``accidental'' cancellation of the one loop quadratic divergences,
the true scale for the naturalness problem might be around 10 TeV
rather than around 1 TeV as it is usually argued. We do not claim, as
Veltman did, that a formula such as eq. (\ref{Veltmanori}) could be
used to compute the Higgs boson's mass. We propose to use Veltman's
relation as a criterion for the naturalness of the model. One could
imagine introducing different cutoffs for the different particles
entering the loop, but even in that case there will be a Veltman type
formula that is either fulfilled or not. If it is fulfilled or even
approximatively fulfilled, it can be interpreted as a sign that the
true scale for the naturalness problem is 10 TeV rather than 1 TeV.

Ultraviolet completions of the models we are proposing have not been
considered. One could imagine having $N$ copies of the standard model
in the case where the mass scale of both scalars is much below the
cutoff scale. Note that if $N \ge 14$, one could consider a cutoff of
the order of the grand unification scale. One could also imagine a
supersymmetric high energy completion in the case where the mass of
one of the scalar doublets is just below the cutoff scale. Another
well studied possibility is that the fundamental scale of physics is
in the TeV region if nature has more than four dimensions
\cite{Arkani-Hamed:1998rs}. In our approach the scale for these new
extra-dimensions might be around 10 TeV and still provide a solution to
the naturalness problem.

\section*{Acknowledgments}
The author is grateful to M. Graesser and S. Su for enlightening
discussions. Furthermore he would like to thank M. Schmaltz for a
useful and interesting communication and P. Bambade for triggering
his interest in the question of the measurement of the Higgs boson
self-coupling.


\begin{thebibliography}{10}
\bibitem{Glashow:1961tr} S.~L.~Glashow,
Nucl.\ Phys.\ {\bf 22}, 579 (1961);
S.~Weinberg,
Phys.\ Rev.\ Lett.\ {\bf 19}, 1264 (1967);
A.~Salam, Elementary Particle Physics, in Proceedings of the 8th Nobel
Symposium, 1968.
\bibitem{Fritzsch:2002jv}
H.~Fritzsch and M.~Gell-Mann, 
``Current algebra: Quarks and what else?,''
 Published in Physics, Proceedings of the XVI International Conference on High Chicago 1972 p.135 (J. D. Jackson, A. Roberts, eds.), 
arXiv:hep-ph/0208010;
H.~Fritzsch, M.~Gell-Mann and H.~Leutwyler,
Phys.\ Lett.\ B {\bf 47}, 365 (1973);
H.~Fritzsch and M.~Gell-Mann, ``Light Cone Current Algebra,''
arXiv:hep-ph/0301127.

\bibitem{Higgs:1964pj}
P.~W.~Higgs,
Phys.\ Lett.\  {\bf 12} (1964) 132; 
Phys.\ Rev.\ Lett.\  {\bf 13} (1964) 508; 
Phys.\ Rev.\  {\bf 145} (1966) 1156; 
F.~Englert and R.~Brout,
Phys.\ Rev.\ Lett.\  {\bf 13} 321 (1964); 
G.~S.~Guralnik, C.~R.~Hagen and T.~W.~Kibble,
Phys.\ Rev.\ Lett.\  {\bf 13} 585 (1964); 
T.~W.~Kibble;
Phys.\ Rev.\  {\bf 155}, 1554 (1967).




\bibitem{Veltman:1981mj}
M.~Veltman,
Acta Phys.\ Polon.\ B {\bf 12}, 437 (1981).




\bibitem{'tHooft:1980xbis} G.~'t~Hooft, in ``Recent Developments In
  Gauge Theories'', Carges\`e 1979, ed.  G.~'t~Hooft et al. Plenum
  Press, New York, 1980, Lecture III, p.135; 
L.~Susskind,
Phys.\ Rev.\ D {\bf 20}, 2619 (1979).

\bibitem{'tHooft:rn} G.~'t Hooft,
Nucl.\ Phys.\ B {\bf 35}, 167 (1971);
Nucl.\ Phys.\ B {\bf 33}, 173 (1971);
G.~'t Hooft and M.~J.~Veltman,
Nucl.\ Phys.\ B {\bf 44}, 189 (1972);
Nucl.\ Phys.\ B {\bf 50}, 318 (1972).
\bibitem{Hill:2002ap}
C.~T.~Hill and E.~H.~Simmons,
arXiv:hep-ph/0203079.

\bibitem{Haber:1984rc}
H.~E.~Haber and G.~L.~Kane,
Phys.\ Rept.\  {\bf 117}, 75 (1985).

\bibitem{Arkani-Hamed:1998rs}
N.~Arkani-Hamed, S.~Dimopoulos and G.~R.~Dvali,
Phys.\ Lett.\ B {\bf 429}, 263 (1998)
[arXiv:hep-ph/9803315];
L.~Randall and R.~Sundrum,
Nucl.\ Phys.\ B {\bf 557}, 79 (1999)
[arXiv:hep-th/9810155];
L.~Randall and R.~Sundrum,
Phys.\ Rev.\ Lett.\  {\bf 83}, 3370 (1999)
[arXiv:hep-ph/9905221].




\bibitem{Georgi:1975tz}
H.~Georgi and A.~Pais,
Phys. Rev.\ D {\bf 12}, 508 (1975);
D.~B.~Kaplan and H.~Georgi,
Phys.\ Lett.\ B {\bf 136}, 183 (1984).






\bibitem{Arkani-Hamed:2002qx} N.~Arkani-Hamed, A.~G.~Cohen, E.~Katz,
A.~E.~Nelson, T.~Gregoire and J.~G.~Wacker,
JHEP {\bf 0208}, 021 (2002)
[arXiv:hep-ph/0206020];
N.~Arkani-Hamed, A.~G.~Cohen, E.~Katz and A.~E.~Nelson,
JHEP {\bf 0207}, 034 (2002)
[arXiv:hep-ph/0206021].


\bibitem{Csaki:2002qg}
C.~Csaki, J.~Hubisz, G.~D.~Kribs, P.~Meade and J.~Terning,
arXiv:hep-ph/0211124,
J.~L.~Hewett, F.~J.~Petriello and T.~G.~Rizzo,
arXiv:hep-ph/0211218;
G.~Burdman, M.~Perelstein and A.~Pierce,
arXiv:hep-ph/0212228;
T.~Han, H.~E.~Logan, B.~McElrath and L.~T.~Wang,
arXiv:hep-ph/0301040.

\bibitem{Decker:gw}
R.~Decker and J.~Pestieau,
Mod.\ Phys.\ Lett.\ A {\bf 4}, 2733 (1989);
E.~Ma,
Phys.\ Rev.\ D {\bf 47}, 2143 (1993)
[arXiv:hep-ph/9209221].


\bibitem{Kolda:2000wi}
C.~F.~Kolda and H.~Murayama,
JHEP {\bf 0007}, 035 (2000)
[arXiv:hep-ph/0003170].


\bibitem{Hagiwara:fs}
K.~Hagiwara {\it et al.}  [Particle Data Group Collaboration],
Phys.\ Rev.\ D {\bf 66}, 010001 (2002).


\bibitem{Pukhov:1999gg}
A.~Pukhov {\it et al.},
``CompHEP: A package for evaluation of Feynman diagrams and integration  over multi-particle phase space. User's manual for version 33,''
hep-ph/9908288.

\bibitem{Bennett:1988xi}
D.~L.~Bennett, H.~B.~Nielsen and I.~Picek,
Phys.\ Lett.\ B {\bf 208}, 275 (1988).



\bibitem{Newton:1993xc}
C.~Newton and T.~T.~Wu,
Z.\ Phys.\ C {\bf 62}, 253 (1994); see also
A.~A.~Andrianov, R.~Rodenberg and N.~V.~Romanenko,
Nuovo Cim.\ A {\bf 108}, 577 (1995)
[arXiv:hep-ph/9408301];
E.~Ma,
Int.\ J.\ Mod.\ Phys.\ A {\bf 16}, 3099 (2001)
[arXiv:hep-ph/0101355].

\bibitem{Calmet:2002rf}
X.~Calmet,
arXiv:hep-ph/0206091.


\end{thebibliography}
\end{document}